\documentclass[final,times,5p]{elsarticle}
\usepackage{epsfig}
\usepackage{graphicx}
\usepackage{amsmath,amssymb}
\usepackage[utf8]{inputenc}\biboptions{sort&compress}
\usepackage{color}
\usepackage[percent]{overpic}
\usepackage{times}
\journal{Physics Letters B}

\begin{document}

\def\pcomma{$\rm^,$}
\def\Mainz{$\rm^a$}
\def\UCLA{$\rm^b$}
\def\Tomsk{$\rm^c$}
\def\Basel{$\rm^d$}
\def\Bonn{$\rm^e$}
\def\Regina{$\rm^f$}
\def\Glasgow{$\rm^g$}
\def\Kent{$\rm^h$}
\def\York{$\rm^i$}
\def\Dubna{$\rm^j$}
\def\Pavia{$\rm^k$}
\def\GWU{$\rm^l$}
\def\LPI{$\rm^m$}
\def\Dalhousie{$\rm^n$}
\def\Halifax{$\rm^o$}
\def\UniPavia{$\rm^p$}
\def\INR{$\rm^q$}
\def\Sackville{$\rm^r$}
\def\Zagreb{$\rm^s$}
\def\Amherst{$\rm^t$}
\def\Jerusalem{$\rm^u$}

\title{Measurement of the beam-helicity asymmetry in photoproduction of $\pi^{0}\eta$ pairs
 on carbon, aluminum, and lead}

\author{
V.~Sokhoyan\Mainz\pcomma$\rm^1$\fntext[fn1]{Corresponding author, email address: sokhoyan@uni-mainz.de},
%V.~Sokhoyan\Mainz\pstar\fntext[fn1]{Corresponding author, email address: sokhoyan@uni-mainz.de},
S.~Prakhov\Mainz\pcomma\UCLA,
A.~Fix\Tomsk,
S.~Abt\Basel,
P.~Achenbach\Mainz,
P.~Adlarson\Mainz,
F.~Afzal\Bonn,
P.~Aguar-Bartolom\'e\Mainz,
Z.~Ahmed\Regina,
K.~Altangerel\Mainz,
J.~R.~M.~Annand\Glasgow,
H.~J.~Arends\Mainz,
K.~Bantawa\Kent,
M.~Bashkanov\York,
R.~Beck\Bonn,
M.~Biroth\Mainz,
N.~S.~Borisov\Dubna,
A.~Braghieri\Pavia,
W.~J.~Briscoe\GWU,
S.~Cherepnya\LPI,
F.~Cividini\Mainz,
C.~Collicott\Dalhousie\pcomma\Halifax,
S.~Costanza\Pavia\pcomma\UniPavia,
A.~Denig\Mainz,
M.~Dieterle\Basel,
E.~J.~Downie\GWU,
P.~Drexler\Mainz,
M.~I.~Ferretti Bondy\Mainz,
L.~V.~Fil'kov\LPI,
S.~Gardner\Glasgow,
S.~Garni\Basel,
D.~I.~Glazier\Glasgow\pcomma\York,
I.~Gorodnov\Dubna,
W.~Gradl\Mainz,
M.~G\"unther\Basel,
G.~M.~Gurevich\INR,
L.~Heijkenskj\"old\Mainz,
D.~Hornidge\Sackville,
G.~M.~Huber\Regina,
A.~K\"aser\Basel,
V.~L.~Kashevarov\Mainz\pcomma\Dubna,
S.~Kay\Regina,
I.~Keshelashvili\Basel,
R.~Kondratiev\INR,
M.~Korolija\Zagreb,
B.~Krusche\Basel,
A.~Lazarev\Dubna,
K.~Livingston\Glasgow,
S.~Lutterer\Basel,
I.~J.~D.~MacGregor\Glasgow,
R.~Macrae\Glasgow,
D.~M.~Manley\Kent,
P.~P.~Martel\Mainz\pcomma\Sackville,
J.~C.~McGeorge\Glasgow,
D.~G.~Middleton\Mainz\pcomma\Sackville,
R.~Miskimen\Amherst,
E.~Mornacchi\Mainz,
C.~Mullen\Glasgow,
A.~Mushkarenkov\Pavia\pcomma\Amherst,
A.~Neganov\Dubna,
A.~Neiser\Mainz,
M.~Oberle\Basel,
M.~Ostrick\Mainz,
P.~B.~Otte\Mainz,
D.~Paudyal\Regina,
P.~Pedroni\Pavia,
A.~Powell\Glasgow,
E.~Rickert\Mainz,
G.~Ron\Jerusalem,
T.~Rostomyan\Basel,
A.~Sarty\Halifax,
C.~Sfienti\Mainz,
K.~Spieker\Bonn,
O.~Steffen\Mainz,
I.~I.~Strakovsky\GWU,
B.~Strandberg\Glasgow,
Th.~Strub\Basel,
I.~Supek\Zagreb,
A.~Thiel\Bonn,
M.~Thiel\Mainz,
A.~Thomas\Mainz,
M.~Unverzagt\Mainz,
Yu.~A.~Usov\Dubna,
S.~Wagner\Mainz,
N.~K.~Walford\Basel,
D.~P.~Watts\York,
D.~Werthm\"uller\York,
J.~Wettig\Mainz,
L.~Witthauer\Basel,
M.~Wolfes\Mainz,
and N.~Zachariou\York
\\
\vspace*{0.1in}
A2 Collaboration at MAMI
\vspace*{0.1in}
}

\address[1]{Institut f\"ur Kernphysik, University of Mainz, D-55099 Mainz,Germany}
\address[2]{University of California Los Angeles, Los Angeles, CA 90095-1547, USA}
\address[3]{Tomsk Polytechnic University, 634034 Tomsk, Russia}
\address[4]{Institut f\"ur Physik, University of Basel, CH-4056 Basel, Switzerland}
\address[5]{Helmholtz-Institut f\"ur Strahlen- und Kernphysik, University of Bonn, D-53115 Bonn, Germany}
\address[6]{University of Regina, Regina, Saskatchewan S4S 0A2, Canada}
\address[7]{SUPA School of Physics and Astronomy, University of Glasgow, Glasgow G12 8QQ, United Kingdom}
\address[8]{Kent State University, Kent, OH 44242-0001, USA}
\address[9]{Department of Physics, University of York, Heslington, York, Y010 5DD, United Kingdom}
\address[10]{Joint Institute for Nuclear Research, 141980 Dubna, Russia}
\address[11]{INFN Sezione di Pavia, I-27100 Pavia, Italy}
\address[12]{The George Washington University, Washington, DC 20052-0001, USA}
\address[13]{Lebedev Physical Institute, 119991 Moscow, Russia}
\address[14]{Dalhousie University, Halifax, Nova Scotia B3H 4R2, Canada}
\address[15]{Saint Mary’s University, Halifax, Nova Scotia B3H 3C3, Canada}
\address[16]{Dipartimento di Fisica, Universit\`a di Pavia, I-27100 Pavia, Italy}
\address[17]{Institute for Nuclear Research, 125047 Moscow, Russia}
\address[18]{Mount Allison University, Sackville, New Brunswick E4L 1E6, Canada}
\address[19]{Rudjer Boskovic Institute, HR-10000 Zagreb, Croatia}
\address[20]{University of Massachusetts, Amherst, MA 01003, USA}
\address[21]{Racah Institute of Physics, Hebrew University of Jerusalem, Jerusalem 91904, Israel}

\date{\today}

\begin{abstract}
 The beam-helicity asymmetry was measured, for the first time, in photoproduction of $\pi^{0}\eta$ pairs
 on carbon, aluminum, and lead, with the A2 experimental setup at MAMI.
 The results are compared to an earlier measurement on a free proton and to the corresponding
 theoretical calculations.
 The Mainz model is used to predict the beam-helicity asymmetry for the nuclear targets.
 The present results indicate that the photoproduction mechanism for $\pi^{0}\eta$ pairs on
 nuclei is similar to photoproduction on a free nucleon. This process is dominated
 by the $D_{33}$ partial wave with the $\eta\Delta(1232)$ intermediate state.
\end{abstract}

\maketitle

\section{Introduction}

In order to understand the spectrum and properties of baryon resonances, significant effort has been made
 during the last decades in studying single- and double-meson
 photoproduction~\cite{PDG,Anisovich2016,BeckThoma2017,CredeRoberts2013,Tiator2011,KlemptRichard2010}.
 In addition to investigating meson photoproduction on a free proton, numerous experiments were
 performed with the aim of understanding meson photoproduction on light and heavy nuclei.
 Meson production on light nuclei, such as the deuteron or helium isotopes, allows one to access
 the baryon resonances produced on the nucleon.
 Photoproduction on heavier targets is well-suited for the understanding of possible modifications
 of hadrons, including baryon resonances, in the nuclear medium.

 One of the most dramatic in-medium effects is the disappearance of the peaks in the second and
 third resonance regions (present in the total photoabsorption on a free proton)
 when heavier targets are used in the experiment~\cite{Frommhold1992,Bianchi1993,Bianchi1994}.
 This observation has not been explained in a model-independent way so far.
 At the same time, it triggered a significant interest in searching for in-medium modifications of
 baryon resonances in exclusive photoproduction channels.
 In the second resonance region, the properties of the $N(1520)3/2^-$ and $N(1535)1/2^-$ resonances
 were studied by using pion and $\eta$ photoproduction on various nuclei
 (see Refs.~\cite{Krusche2011,KruscheReview2012,PPN-BK} for overview), and numerous studies showed that these
 resonances are not strongly modified in the nuclear medium~\cite{Krusche2004,PPN-BK,PLB-Robig,PLB-Yor,NP-Yam}.
 Possible modifications of baryon resonances in the third resonance region have not been investigated
 in such detail, and the photoproduction mechanisms of the $\Delta(1700)3/2^-$ and
 $\Delta(1940)3/2^-$ resonances on heavy nuclei have not been studied thus far.
 Photoproduction of $\pi^0\eta$ pairs on nuclei is well suited for accessing the properties of these resonances due to a selective identification of contributing resonances and their decay modes,
 compared to widely investigated $2\pi^0$ photoproduction, where the reaction dynamics is much more complicated
 (see e.g. Refs.~\cite{Dieterle2015,Sokhoyan2015EPJA,Sokhoyan2015PLB,Thiel2015,Oberle2013,Zehr2012,p2pi0mamic,Thoma2008,Sarantsev2008,Strauch2005,Assafiri2003}).
 For the incoming-photon energy range from the production threshold up to $E_{\gamma} = 1.5$~GeV,
 numerous analyses indicate the dominance of the $D_{33}$ partial
 wave~\cite{Doering2006,Mainz_2008,Mainz_2010,Kashev_pi0eta_prod,Kashev_pi0eta_asym,A2_pi0etap_2018,CBELSA_pi0etap_2008,CBELSA_pi0etap_2010,CBELSA_pi0etap_2014},
 which couples strongly to the $\Delta(1700)3/2^-$ resonance close to the production threshold
 and to the $\Delta(1940)3/2^-$ at higher energies.
 Another advantage of $\pi^0\eta$ photoproduction is that the $\eta$ meson, serving as an isospin filter,
% allows access to transitions between either two $N^*$ or two $\Delta$ resonances,
 allows access to transitions between two different $N^*$ or two different $\Delta$ resonances
 (from a heavier to a lighter one),
 thus providing additional selectivity to the investigated decay mode in case of sequential decays
 with independent emission of the two mesons.

 Photoproduction of $\pi^0\eta$ pairs has been extensively studied on a free proton,
 with various angular differential cross sections and distributions for polarization observables reported earlier in Refs.~\cite{CBELSA_pi0etap_2008,CBELSA_pi0etap_2010,CBELSA_pi0etap_2014,Kashev_pi0eta_prod,Kashev_pi0eta_asym,Mainz_2010,A2_pi0etap_2018,GRAAL_2008,LNS2006}.
 The A2 collaboration recently reported the unpolarized cross sections and the beam-helicity asymmetry
 for photoproduction of $\pi^{0}\eta$ pairs on the deuteron and on helium nuclei~\cite{Kaeser2016}
 and the helicity-dependent cross sections for photoproduction
 of $\pi^{0}\eta$ pairs on quasi-free protons and neutrons~\cite{Kaeser2018}.
 The results on the deuteron confirmed that the $\pi^{0}\eta$ production mechanism on the proton and
 neutron is dominated by the $D_{33}$ partial wave with the $\eta\Delta(1232)$ intermediate state.

  In the present work, photoproduction of $\pi^{0}\eta$ pairs was investigated for the first time
 by using circularly polarized photons incident on heavier nuclear targets (carbon, aluminum, and lead).
 The main goal of the experiment was to test whether the mechanism for $\pi^{0}\eta$ photoproduction
 on heavy nuclei is also dominated by the $D_{33}$ partial wave.
 Such a study was inspired by results from Ref.~\cite{Kaeser2016}, which showed that, although the overall
 rate for $\pi^{0}\eta$ photoproduction is reduced significantly on the deuteron
 due to final-state interactions (FSI), the beam-helicity asymmetry, $I^{\odot}$, remains practically unchanged.
 This feature allows one to use $I^{\odot}$ measured in $\pi^{0}\eta$ photoproduction for investigating
 possible changes in the production mechanisms of the $D_{33}$ wave on heavier nuclei,
 presumably not being strongly influenced by the FSI-related effects.

As shown in Ref.~\cite{Kashev_pi0eta_asym}, the asymmetry $I^{\odot}$ originating from the $D_{33}$ partial wave should have a specific shape
 \begin{equation}\label{sin}
 I^\odot(\Phi_\pi) = A_1 \sin \Phi_\pi + A_2 \sin 2\Phi_\pi
\end{equation}
% as a function of the angle $\Phi_{\pi}$, which is the azimuthal pion angle in the $\pi N$ rest frame
% with respect to the plane determined by the momenta of the $\pi N$ system and
% the incident photon in the overall center-of-mass (c.m.) frame.
as a function of the azimuthal angle $\Phi_{\pi}$, which is the angle between the pion in the $\pi N$
 rest frame and the plane determined by the momentum of the $\pi N$ system in the overall center-of-mass
 (c.m.) frame and the incident photon in the same c.m. frame.
 Then the first term in Eq.~(\ref{sin}) is determined solely by the $D_{33}$ wave, and the second by its interference with other waves.
 The first experimental measurement of the beam-helicity asymmetry, reported in Ref.~\cite{Kashev_pi0eta_asym},
 did reveal that its shape was similar to sinusoidal, especially in the energy range close to the production threshold.
 Such an observation confirmed the strong dominance of the $D_{33}$ partial wave
 in $\gamma N\to\pi^0\eta N$, independently of the earlier result based on the analysis on other unpolarized
 observables~\cite{Kashev_pi0eta_prod}. This feature makes the $I^\odot$ observable well suited for studying
 the behavior of the elementary amplitude in a nuclear environment, as possible significant changes in the partial-wave
 structure of a single-nucleon amplitude in a nucleus will lead to noticeable deviations from the $I^\odot(\Phi_\pi)$
 dependence observed on a free nucleon. The most likely modification for the $D_{33}$ states in the nuclear environment (beyond FSI) would be a suppression of intensity or an increase of their width due to the presence of different inelastic mechanisms. In this case, the relative contribution of other terms could be changed, resulting in a different energy dependence
 of the beam-helicity asymmetry and in
% a distortion of the $I^\odot(\Phi_\pi)$ dependence from the shape determined by Eq.~(\ref{sin}).
 a deviation of the $I^\odot(\Phi_\pi)$ dependence from the shape determined by Eq.~(\ref{sin}).

 In this work, the results obtained for the beam-helicity asymmetry on the three nuclear targets
 were compared to the earlier A2 results on a free proton~\cite{A2_pi0etap_2018} and
% to the corresponding theoretical calculations with the latest~\cite{A2_pi0etap_2018}
% and the earlier the Mainz model~\cite{Kashev_pi0eta_asym}.
 to the corresponding theoretical calculations with the latest version of
 the Mainz model~\cite{A2_pi0etap_2018}.
 The Mainz model was initially developed for the analysis of three-body final states,
 especially aiming for understanding the features of $\gamma N\to \pi\pi N$ and
 $\gamma N\to \pi^0\eta N$~\cite{Mainz_2005,Mainz_2008,Mainz_2010,Mainz_2011,Mainz_2013}.
 The model parameters were adjusted by simultaneously fitting various experimental distributions
 for observables sensitive to the reaction dynamics, paying particular attention to the analysis
 of specific angular distributions. Based on the parameters adjusted for free nucleons, this model can also make predictions for quasi-free nucleons and heavier nuclei.
 In this work, the Mainz model~\cite{A2_pi0etap_2018} was used to predict the change in
 the beam-helicity asymmetry in the transition from a free nucleon to the three nuclear targets
 used in the present experiment.

 There is also a simultaneous partial-wave analysis (PWA) of available photoproduction data
 by the Bonn-Gatchina (BnGa) group~\cite{BGPWA},
 the results of which for the $\gamma p\to \pi^0\eta p$ data from CBELSA/TAPS
 were earlier reported in Refs.~\cite{CBELSA_pi0etap_2008,CBELSA_pi0etap_2010,CBELSA_pi0etap_2014}.
 %Note that no beam-helicity asymmetry data were included in the BnGa PWA.
 Note that no beam-helicity asymmetry data are included in the BnGa PWA.
 As shown in Ref.~\cite{A2_pi0etap_2018}, the solutions of the BnGa PWA for the beam-helicity asymmetry on a free proton are in good agreement with the A2 data for the lower energy range, and then the discrepancy increases with energy.
There are no solutions by the BnGa PWA for the nuclear targets used in our experiment.

\section{Experimental setup}
\label{sec:Setup}

Photoproduction of $\pi^{0}\eta$ pairs on nuclear targets was measured at the Mainz Microtron (MAMI)~\cite{MAMI,MAMIC}, using an energy-tagged bremsstrahlung photon beam. The energies of the incident photons were analyzed up to 1400~MeV, by detecting the postbremsstrahlung electrons in the Glasgow tagged-photon spectrometer (Glasgow tagger)~\cite{TAGGER,TAGGER1,TAGGER2}.
The uncertainty of $\pm 2$~MeV in the energy of the tagged photons is mostly determined by the segmentation of the  focal-plane detector of the Glasgow tagger in combination with the energy of the MAMI electron beam (more details are given in Ref.~\cite{TAGGER2}).

The final-state particles were detected by using the Crystal Ball (CB)~\cite{CB} as a central calorimeter and TAPS~\cite{TAPS,TAPS2} as a forward calorimeter.
The CB detector consists of 672 NaI(Tl) crystals covering polar angles from $20^{\circ}$ to $150^{\circ}$. The TAPS
calorimeter consists of 366 BaF$_2$ crystals covering polar angles from $4^{\circ}$ to $20^{\circ}$ and 72 PbWO$_{4}$ crystals with angular coverage from $1^{\circ}$ to $4^{\circ}$.
Both CB and TAPS calorimeters have full azimuthal coverage. More information on the energy and angular resolution of the CB and TAPS is provided in Refs.~\cite{etamamic,slopemamic}.

 The target, located in the center of the CB, was surrounded by a Particle IDentification
 (PID) detector~\cite{PID}, consisting of 24 scintillator bars, and by two Multiwire Proportional Chambers (MWPCs),
 serving for identification and tracking of charged particles. In the TAPS region, plastic veto detectors
 were placed in front of the BaF$_2$ and PbWO$_{4}$ crystals.

 The present measurements were conducted with a 1557-MeV beam of longitudinally polarized electrons from the Mainz Microtron, MAMI-C~\cite{MAMIC}.
 Circularly-polarized bremsstrahlung photons, incident on the solid targets, were produced by the beam electrons
 in a 10-$\mu$m radiator made of iron and cobalt alloy and collimated by a 2.5-mm-diameter Pb collimator.
 Experimental data were measured with carbon, aluminum, and lead targets with thickness
 of 20, 8, and 0.5 mm, respectively.
 The photon degree of polarization was determined as~\cite{Olsen_1959}
\begin{equation} \label{pgamma}
P_\gamma =P_{e^-} \frac{4x-x^2}{4-4x+3x^2}~,
\end{equation}
  where $P_{e^-}$ is the electron degree of polarization,
 and $x=E_\gamma/E_{e^-}$ is the ratio of a bremsstrahlung-photon energy to the energy of the electron beam
 from MAMI. In the present measurements, the averaged magnitude
 of $P_{e^-}$ was 0.745, 0.705, and 0.715 for the carbon, aluminum, and lead targets,
 respectively.

 The experimental trigger first required the total energy deposited in the CB to exceed $\sim$320~MeV for the aluminum and lead targets and $\sim$350~MeV for the carbon target. Then the number of so-called hardware clusters in the CB and TAPS together (multiplicity trigger) had to be two or more.

\section{Data handling}
\label{sec:Data}

 Events from photoproduction of $\pi^0\eta$ pairs on nuclei were searched for in
 the four-photon final state produced by $\pi^0\to \gamma\gamma$ and $\eta\to \gamma\gamma$ decays.
 The reaction candidates were extracted from events with four or
% five clusters reconstructed in the CB and TAPS together by a software analysis.
 five clusters reconstructed in the CB and TAPS together with a software analysis. Four-cluster events were analyzed by assuming that only four final-state photons had been detected, and five-cluster events by assuming that the recoil nucleon
 (proton or neutron) from a nucleus had been detected as well.
% new
 The separation of $\pi^0\eta$ events produced on protons or neutrons was not used in the analysis
 because it is impossible for four-cluster events.

 Similar to the analysis of the data with a hydrogen target~\cite{A2_pi0etap_2018},
 kinematic fitting was used to select event candidates and to reconstruct
 the reaction kinematics. Details of the kinematic-fit parametrization of
 the detector information and resolutions are given in Ref.~\cite{slopemamic}.
 Unlike for the free-proton case, the missing mass
 of the four-photon final state was used in the reaction hypothesis.
% This missing mass was calculated by assuming that the target particle
% has the nucleon mass.
 This missing mass was calculated by assuming that the target particle
 has the nucleon mass and zero momentum. To identify events from the $\pi^0\eta$ photoproduction,
 the $\gamma N\to \pi^0\gamma\gamma X\to 4\gamma X$ hypothesis, using only the $\pi^0$-mass constraint
 on the two-photon invariant mass, was tested, and the events that satisfied this hypothesis
 with probability greater than 2\% were selected for further analysis.
 Then, for the selected events, a peak from $\eta\to \gamma\gamma$ decays can be
 seen in the invariant-mass distribution of the two photons that are not
 from the $\pi^0$ decay.
 The background under the $\eta\to \gamma\gamma$ peak comes mostly from
 misidentification of clusters from neutrons and charged particles with photons.
 At the same time, the $\eta\to \gamma\gamma$ peak itself can include the background
 from $\gamma n\to \pi^-\eta p$ and $\gamma p\to \pi^+\eta n$ events, when the invariant mass
 of two clusters either from $\pi^-p$ or from $\pi^+n$ is close to the $\pi^0$ mass.
 Based on Monte Carlo (MC) simulations of these two processes,
 compared to $\gamma N\to \pi^0\eta N\to 4\gamma N$,
 such background contributes less than 5\% in the signal peak.

\begin{figure*}
\includegraphics[width=0.99\textwidth]{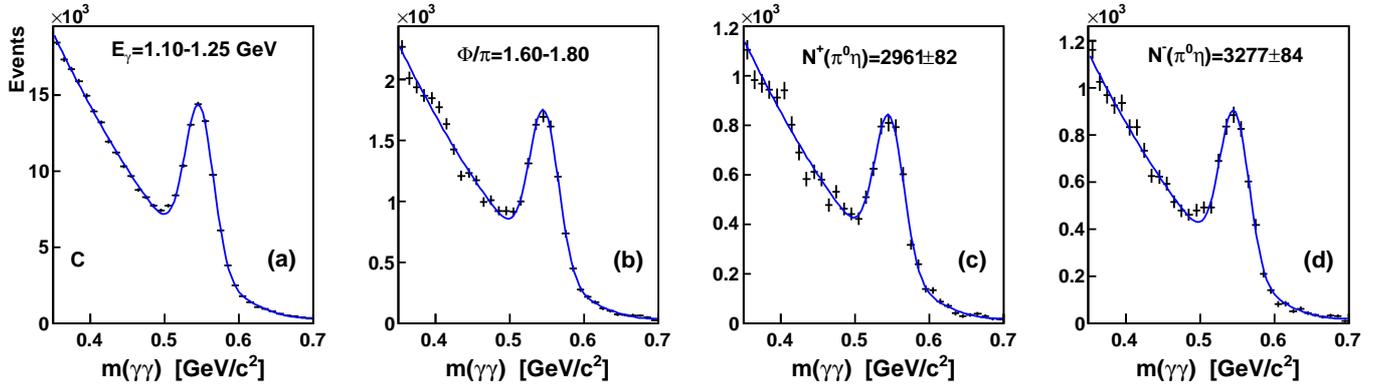}
\caption{
 (a) Invariant-mass distribution of the two photons not from the $\pi^0$ decay
 for $\gamma N\to \pi^0\gamma\gamma X\to 4\gamma X$ events produced on
 the carbon target within $E_{\gamma}=1.10-1.25$~GeV; (b) same as (a) but
 within $\Phi/\pi=1.6-1.8$; (c) same as (b) but for helicity (+) only;
 (d) same as (b) but for helicity (-) only.
 Fits to the distributions, which were made with the sum
 of a Gaussian for the $\eta \to \gamma\gamma$ peak and
 a polynomial of order 4 for the background, are shown by the solid blue lines.
}
 \label{fig:pi0eta_c_mgg_fit}
\end{figure*}
\begin{figure*}
\includegraphics[width=0.99\textwidth]{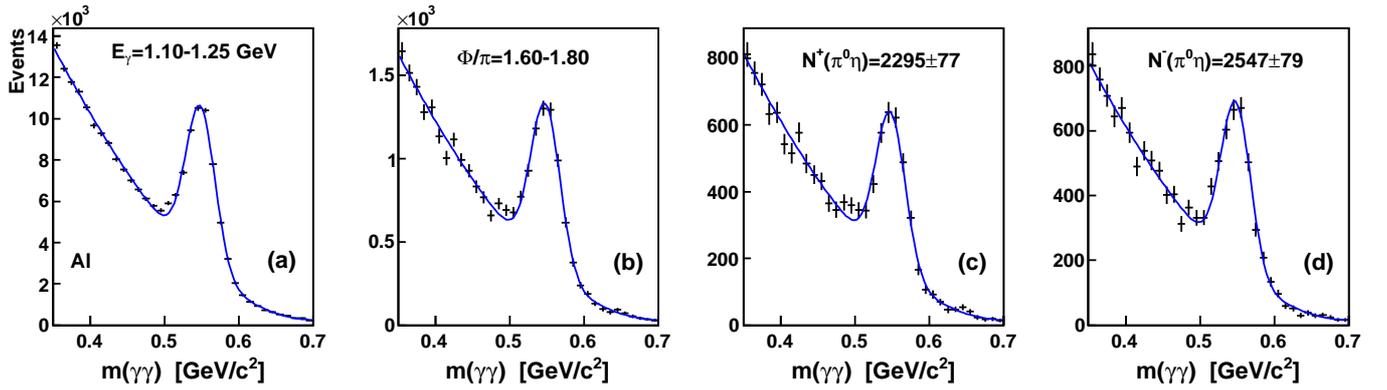}
\caption{
  Same as Fig.~\protect\ref{fig:pi0eta_c_mgg_fit} but for the aluminum target.
}
 \label{fig:pi0eta_al_mgg_fit}
\end{figure*}
\begin{figure*}
\includegraphics[width=0.99\textwidth]{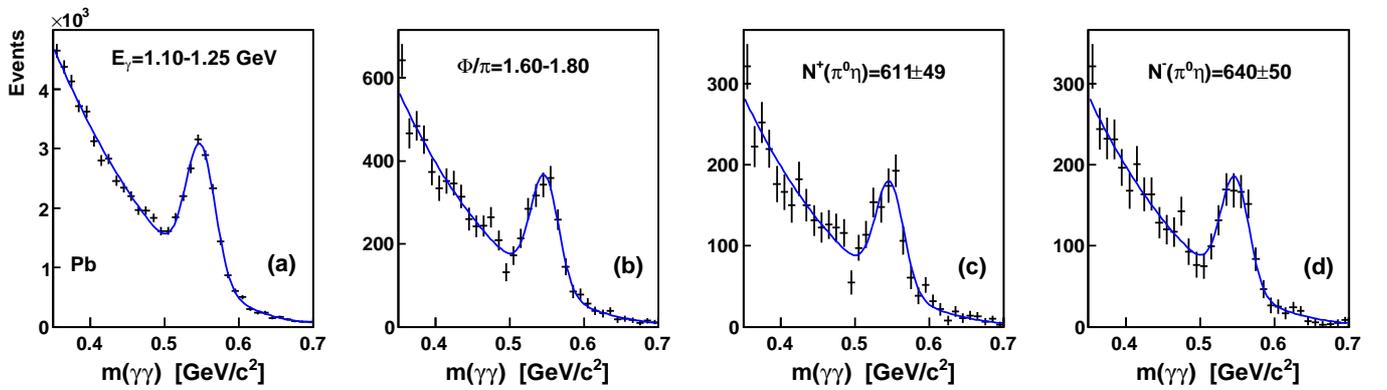}
\caption{
  Same as Fig.~\protect\ref{fig:pi0eta_c_mgg_fit} but for the lead target.
}
 \label{fig:pi0eta_pb_mgg_fit}
\end{figure*}

 To calculate $I^\odot$, the transformation to the c.m.\ frame was
 made by assuming that the target particle has the nucleon mass and zero momentum,
  which is similar to the hypothesis used in the kinematic fit, and the mass of the recoil
 particle $X$ was taken as the missing mass.
 Then the asymmetry due to the photon-beam helicity can be defined as a function
 of the angle $\Phi$ between the production plane and the reaction plane.
 For a three-particle final state, all those three particles in the c.m.\ frame lie in
 the same plane, which is typically called the production plane.
 A reaction plane is typically determined by the beam particle and one of the
 three final-state particles in the c.m.\ frame.
 In the previous A2 measurements with a hydrogen target~\cite{Kashev_pi0eta_asym,A2_pi0etap_2018},
 the reaction plane was determined by the vector product of the momenta of the $\eta$ meson and
 the incident photon. Such a choice is more informative when the production is dominated by
 the $\eta \Delta(1232)$ intermediate state, which was observed in this energy range.
 Then the production plane rotates around the back-to-back directions of
 $\eta$ and $\Delta(1232)$. The orientation of the production and the reaction plane was then
 chosen in such a way that the angle $\Phi$ had to be identical to the angle $\Phi_{\pi}$ used in Eq.~(\ref{sin}).
Because the purpose of this work was to test whether the same production
 mechanisms dominate in heavier nuclei, the reaction plane was determined similar to the previous analyses.

 Experimentally, the beam-helicity asymmetry $I^\odot(\Phi)$ can be measured as
\begin{equation} \label{asym}
  I^\odot(\Phi) = \frac{d\sigma^+ - d\sigma^-}{d\sigma^+ + d\sigma^-} =
  \frac{1}{P_\gamma}\frac{N^+ - N^-}{N^+ + N^-}~,
\end{equation}
 where $d\sigma^{\pm}$ are the differential cross sections as a function of $\Phi$ for each of the two helicity states
 of the incident photon, $P_\gamma$ is the degree of circular polarization of the photon,
 and $N^{\pm}$ are the number of events produced at the angle $\Phi$ for the two helicity states.
 It was checked with the MC events weighted with the BnGa PWA polarized amplitude that the experimental acceptance as a function of $\Phi$ is identical for $N^+$ and $N^-$,
 and its impact on $N^{+/-}(\Phi)$ is canceled in the ratio $(N^+ - N^-)/(N^+ + N^-)$.

 To measure $I^\odot(\Phi)$, all $\gamma N\to \pi^0\gamma\gamma X\to 4\gamma X$ events selected
 for each target and beam helicity
 were divided into 3 energy (150 MeV wide) intervals, with 10 angular bins in $\Phi$. To compare the results on nuclear targets with hydrogen, the data of 50-MeV-wide bins from Ref.~\cite{A2_pi0etap_2018}
 were combined in the corresponding 150-MeV-wide bins,
 with similar modifications made for the free-proton predictions.
% with the Mainz model~\cite{Mainz_2010,A2_pi0etap_2018} and BnGa PWA~\cite{CBELSA_pi0etap_2014}.
 The invariant-mass distributions of the two photons not from the $\pi^0$ decay were used
 to determine $N^+(\Phi)$ and $N^-(\Phi)$ in each bin.
 These distributions are illustrated in Figs.~\ref{fig:pi0eta_c_mgg_fit},~\ref{fig:pi0eta_al_mgg_fit},
 and~\ref{fig:pi0eta_pb_mgg_fit} for the carbon, aluminum, and lead targets, respectively.
 For a better comparison of the experimental statistics available for each target,
 the invariant-mass distributions are shown for the same bin, with the incident-photon energy range
 $1.10<E_{\gamma}<1.25$~GeV and the angular range $1.6<\Phi/\pi<1.8$.
 To measure the number of $\gamma N\to \pi^0\eta X\to 4\gamma X$ events, the invariant-mass distributions
 were fitted with the sum of a Gaussian for the $\eta \to \gamma\gamma$ peak and
 a polynomial of order 4 for the background. Fits to the distributions shown in plots (a),
 obtained for all events in the given energy range,
 were used to determine the initial parameters of the fits to the distributions in plots (b),
 additionally restricted with events only within $1.6<\Phi/\pi<1.8$.
 Plots (c) and (d) divide the events from plots (b) based on their helicity state.
 In the fits to these $m(\gamma\gamma)$ distributions, the parameters describing the Gaussian and
 the polynomial shape were fixed to the fit results from plot (b),
 and only the weights of the two functions were free parameters of the fit.
 For data with low statistics as for the lead target, such an approach provides more reliable results
 for measuring $N^+(\Phi)$, $N^-(\Phi)$, and the corresponding $I^\odot(\Phi)$, compared to completely
 independent fits to the final $m(\gamma\gamma)$ distributions.

\section{Mainz model for nuclei}
\label{sec:Model}
 The theoretical calculations for photoproduction of $\pi^0\eta$ pairs on heavy nuclei were made
 within the Mainz isobar model~\cite{Mainz_2010}, revised recently for the analysis
 of the latest A2 data on $\gamma p\to \pi^0\eta p$~\cite{A2_pi0etap_2018}.
 Within this model, the $\gamma N\to\pi^0\eta N$ amplitude consists of the three main terms:
 the resonant sector, the Born amplitudes, and additional background contributions.
 The first two terms are basically similar to those from an earlier version of the Mainz
 model~\cite{Mainz_2010} used to describe the first A2 data
 on $\gamma p\to \pi^0\eta p$~\cite{Kashev_pi0eta_prod,Kashev_pi0eta_asym}.
 The Born amplitudes contain the diagrams with the nucleon and $N(1535)$ poles in
 the $s$- and $u$-channels, without any parameters that could be adjusted from fitting
 to experimental data. Direct calculations show that the Born amplitudes contribute only a small fraction
 ($\sim 1\% - 2\%$) to the total cross section.
 In addition, to improve the quality of data description in Ref.~\cite{A2_pi0etap_2018},
 artificial background terms were included into the partial waves with $J\leq 5/2$.
 The major constraint of the model is that these terms should be small in magnitude and have
 smooth energy dependence. The resonant part of the amplitude contains
 four $\Delta$-type resonances rated by four stars in the Review of Particle Physics (RPP)~\cite{PDG}:
 $\Delta(1700)3/2^-$, $\Delta(1905)5/2^+$, $\Delta(1920)3/2^+$, and $\Delta(1940)3/2^-$.
 According to the analysis in Ref.~\cite{A2_pi0etap_2018}, the contribution from other resonances
 in the present energy range is small and effectively contained in
 the background term. The parameters obtained earlier for the four $\Delta$ resonances
 are given in Ref.~\cite{A2_pi0etap_2018}.

 To calculate $\pi\eta$ photoproduction on nuclei, a spectator model was adopted in connection with the closure
 relation $\sum\limits_f|f\rangle\langle f|=1$ for the sum over states of the residual nuclear system.
% For simplicity, the residual system of $A-1$ nucleons was treated as if it were in its ground state.
 For simplicity, the residual nuclear system in the reaction kinematics was treated as if it were a bound system
 of $A-1$ nucleons (further denoted as a nucleus $A_f$) with the mass $M_A-M_N$, where $M_A$ is the target-nucleus
 mass and $M_N$ is the nucleon mass.
 Then, the nuclear cross section in the overall c.m.\ frame,
 corresponding to the helicity component $\pm 1$ of the incident photon, can be presented
 in terms of the square of the spin-averaged single-nucleon amplitude $t^{\pm}_{\gamma N}$ as
\begin{eqnarray}\label{XsectTheory}
&&\frac{d\sigma^{\pm}_{\gamma A}}{d\Phi_{\pi}}
=2\pi\int\,{\cal K}
f_\pi(T_\pi)\,f_\eta(T_\eta)\,\rho_A(p)\overline{|t^{\pm}_{\gamma N}|^2}\\
&&\phantom{xx}d\omega_{\pi N}\,d\cos\theta^*_{\pi N}\,
d\omega_{\eta A_f}\,d\Omega^*_{\eta A_f}\,d\cos\Theta_{\eta A_f}\,,\nonumber
\end{eqnarray}
where the kinematic phase-space factor has the form
\begin{equation}\label{K}
{\cal K}=\frac{1}{(2\pi)^8}\,\frac{E_i\,E_f M_Nq^*_{\pi N}\,p^*_{\eta A_f}\,P_{\eta A_f}}{8W^2E_\gamma}\,,
\end{equation}
 with $W$, $E_\gamma$, $E_i$, and $E_f$ being the total c.m.\ energy, the energies of the incident photon,
 and of the initial and the final nucleus, respectively.
 The notations $\omega_{\pi N}$ and $\omega_{\eta A_f}$ are used for the invariant masses of the $\pi N$
 and $\eta A_f$ systems, with the 3-momenta $q^*_{\pi N}$ and $p^*_{\eta A_f}$ and spherical angles
 $\Omega^*_{\pi N}=\{\theta^*_{\pi N},\,\Phi_{\pi}\}$ and $\Omega^*_{\eta A_f}$ in the corresponding rest frames.
 $P_{\eta A_f}$ is the 3-momentum of the $\eta A_f$ system in the overall c.m.\ frame.

In Eq.\,(\ref{XsectTheory}), the factors $f_\pi(T_\pi)$ and $f_\eta(T_\eta)$, which depend on
 the kinetic energies of the particles, are introduced to take into account absorption of
 the produced mesons. Their calculation assumes a square-well approximation for the $\pi$-nucleus and
 $\eta$-nucleus optical potential. Then these attenuation factors can be obtained in
 a simple analytic form (see, for example, Ref.~\cite{Laget}) as
\begin{eqnarray}
&&f_{\alpha}(T_\alpha)=\frac{3\lambda_\alpha}{4R}\Big[
1-\\
&&\phantom{xx}-\frac{\lambda_\alpha^2}{2R^2}\Big\{
1-\Big( 1+\frac{2R}{\lambda_\alpha}\Big)\,e^{-2R/\lambda_\alpha}\Big\}\Big]\,,
\quad \alpha=\pi,\eta\,,\nonumber
\end{eqnarray}
where $R$ and $\lambda_\alpha$ are the square-well nuclear radius and the mean free path of
 the meson $\alpha$ in nuclear matter, respectively. However, it is worth noting that
 the direct calculation shows an insignificant influence of the attenuation factors $f_\pi$
 and $f_\eta$ on the observable $I^\odot$. This result can primarily be explained by the smooth energy dependence
 of those factors, which leads to the significant cancellation of absorption effects in the ratio (\ref{asym}).
\begin{figure*}
      \begin{overpic}[width=0.95\textwidth
      % ,grid,tics=1
]
{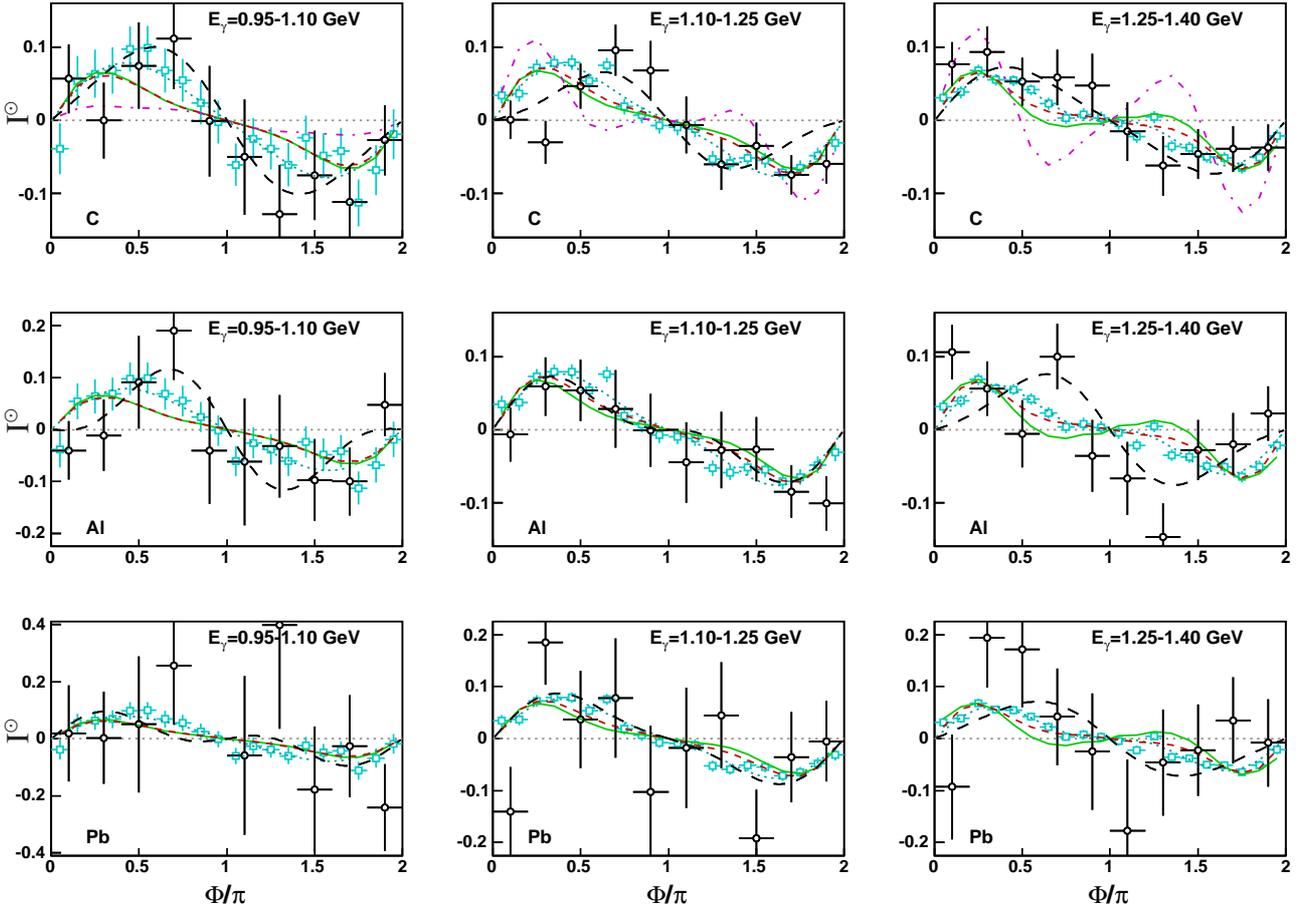}
\put(2.5, 13){\begin{large}\sffamily{\rotatebox{90}{$\rm I^\odot$}}\end{large}}
\put(2.5, 36.6){\begin{large}\sffamily{\rotatebox{90}{$\rm I^\odot$}}\end{large}}
\put(2.5, 60){\begin{large}\sffamily{\rotatebox{90}{$\rm I^\odot$}}\end{large}}

\end{overpic}
\caption{
 Comparison of the present results for the beam-helicity asymmetry $I^\odot$ on the nuclear
 targets (open circles) to the data points obtained on a free proton (cyan open
 squares)~\protect\cite{A2_pi0etap_2018}, to the Mainz model~\protect\cite{A2_pi0etap_2018}
 for a free proton (red dashed line) and for the nuclear targets (solid green line).
 The change in the Mainz model after removing the $D_{33}$ contribution is
 illustrated for the carbon target (magenta dash-dotted line).
 Fits to the experimental data points with
 Eq.~(\protect\ref{sin}) are shown with the long-dashed black line for the nuclear targets and
 with the cyan dotted line for the hydrogen target.
 The top, middle, and bottom rows show results for the carbon, aluminum, and lead targets, respectively.
 A data point $I^\odot=-0.65\pm0.30$ of the angular bin $0.8<\Phi/\pi<1.0$ for Pb at $0.95<E_\gamma<1.10$~GeV
 is out of the plotted range.
}
 \label{fig:Io_asym_solid_3x3}
\end{figure*}

 An important component in Eq.\,(\ref{XsectTheory}) is the function $\rho_A(p)$, which describes
 the distribution of the initial bound nucleon in terms of its momentum $p$.
 In the actual calculation for $^{12}$C, a harmonic oscillator potential is used, yielding
 a well-known form for the nucleon-momentum distribution,
\begin{equation}
\rho_{^{12}\mbox\small{C}}(p)=8\pi\,\sqrt{\pi} r_0^3\left(N_s+\frac23N_p\,p^2r_0^2\right)
e^{-(pr_0)^2}\,,
\end{equation}
where $N_s=2$ and $N_p=4$ are the number of protons (neutrons) on the $s$- and $p$-shell of $^{12}$C.
The value $r_0=1.64$\,fm was used for the oscillator parameter.
Calculations for $^{27}$Al and $^{208}$Pb adopted the momentum distributions from Ref.\,\cite{Ryckebusch},
 where the effects of short-range correlations were taken into account as well.

\section{Results and discussion}
\label{sec:Results}

 The results obtained in this work for the beam-helicity asymmetry $I^\odot$ on the three nuclear targets
 are shown in Fig.~\ref{fig:Io_asym_solid_3x3}.
 The results are compared to the previous A2 measurement on a free proton~\cite{A2_pi0etap_2018},
 which for convenience are plotted in with a finer binning in $\Phi$,
 and to the corresponding calculations with the Mainz model~\cite{A2_pi0etap_2018},
 which was also used to predict the beam-helicity asymmetry for the three nuclear targets.
% The adjustment of the latest model involved a much larger number of
% experimental observables, compared to the earlier model.
% As shown in Fig.~\ref{fig:Io_asym_solid_3x3}, the earlier Mainz model gives a better agreement
% with the the free-proton data at lower energies and the latest model at higher energies.
% The solutions for the BnGa PWA~\cite{CBELSA_pi0etap_2014}
% are not plotted because of their poor agreement with the free-proton data at higher energies.
% As also shown in Fig.~\ref{fig:Io_asym_solid_3x3}, the calculations made with the latest Mainz model
 As shown in Fig.~\ref{fig:Io_asym_solid_3x3}, the calculations made with the Mainz model
 predict a very similar $I^\odot(\Phi)$ dependence for heavy nuclei and a free nucleon,
 especially near the production threshold. This demonstrates that the direct comparison of the present
 heavy-nuclei data with earlier measurements and calculations on a free nucleon is quite fair.
 The data points obtained for the nuclear targets are for the most part in agreement within the error bars with the data points for a free proton.
 Larger uncertainties for the lowest-energy bin, especially for the lead target,
 make difficult the visual comparison of the $I^\odot(\Phi)$ dependences from different targets and do not allow the firm conclusion on their similarity.

 The uncertainties in the $I^\odot(\Phi)$ data points obtained for the nuclear targets are based on the uncertainties in $N^+(\Phi)$ and $N^-(\Phi)$ extracted from the parameter errors of the fits to the corresponding $m(\gamma\gamma)$ distributions. The latter uncertainties
 depend on both the number of $\pi^0\eta$ events detected and the level of background
 under the $\eta\to\gamma\gamma$ peak.
 The uncertainties in the $I^\odot(\Phi)$ data points for a free proton are simply statistical
 (see Ref.~\cite{A2_pi0etap_2018} for more details).

\begin{figure}
\includegraphics[width=0.45\textwidth]{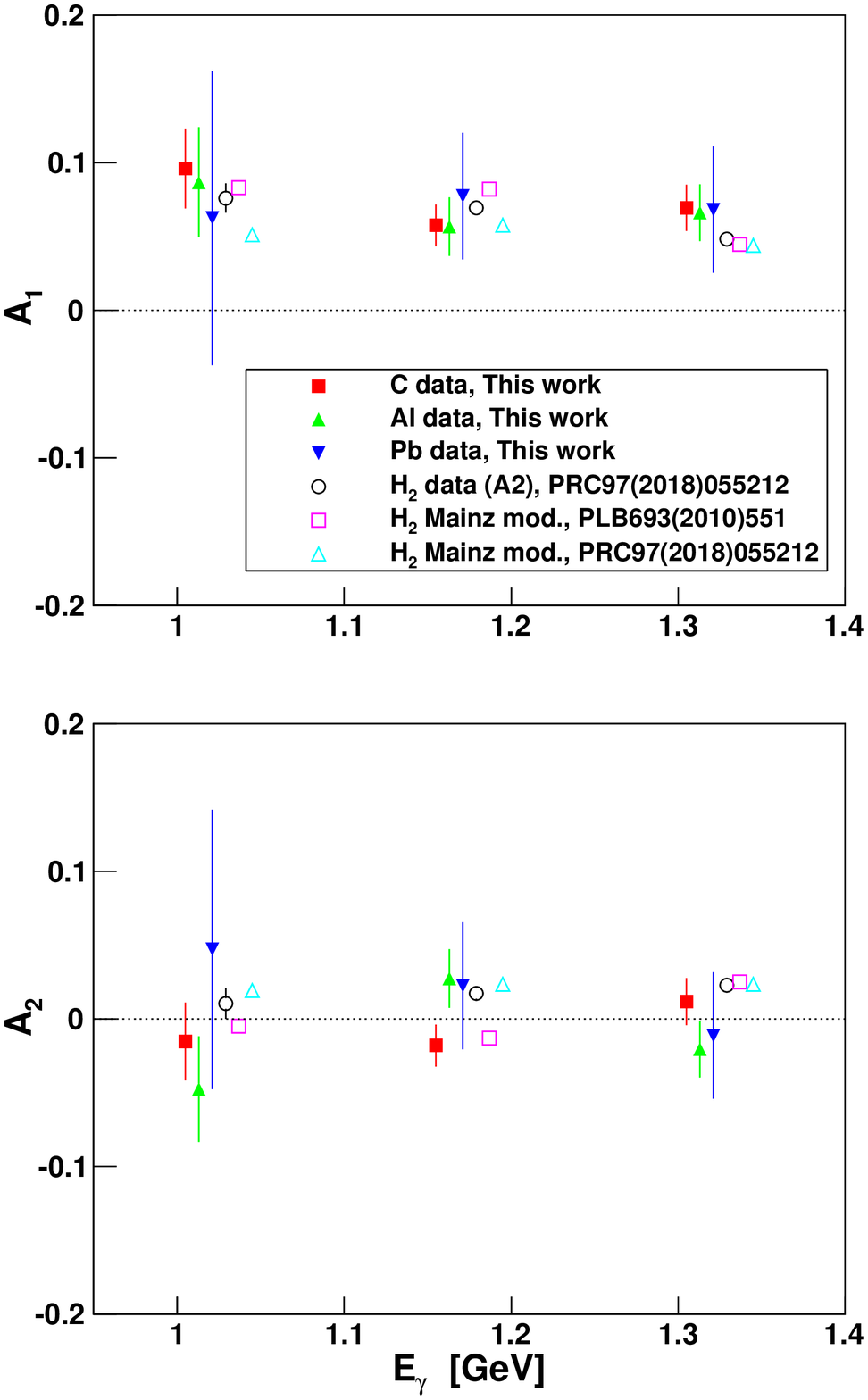}
\caption{
 Comparison of the coefficients $A_1$ and $A_2$ of Eq.~(\protect\ref{sin}) used to fit
 to the data points obtained for the three nuclear targets,
 to the free-proton data ~\protect\cite{A2_pi0etap_2018}, and to the corresponding predictions
 with the Mainz model~\protect\cite{Kashev_pi0eta_asym,A2_pi0etap_2018}.
}
 \label{fig:pi0eta_solid_coef}
\end{figure}

 For data points with large error bars, a better comparison of magnitudes and shapes of
 the $I^\odot(\Phi)$ dependences can be made by fitting them with
 function from Eq.~(\ref{sin}). Those fits obtained for the three nuclear targets and the free-proton data are shown in Fig.~\ref{fig:Io_asym_solid_3x3}; they indicate a similar magnitude for their $I^\odot(\Phi)$ dependences, but with some shift in $\Phi$ sometimes.
 Numerically, the results of those fits to different data can be compared via the values obtained
 for the coefficients $A_1$ and $A_2$ of Eq.~(\ref{sin}).
 In Fig.~\ref{fig:pi0eta_solid_coef}, these coefficients for the three nuclear targets are compared
 to each other and to the coefficients obtained for the free-proton data and to the corresponding
 predictions with the Mainz model~\cite{Kashev_pi0eta_asym,A2_pi0etap_2018}. The coefficients for the three nuclear targets are very similar to the values obtained for the free-proton prediction with the Mainz model and, therefore, are not shown in Fig.~\ref{fig:pi0eta_solid_coef}.

 As shown in Fig.~\ref{fig:pi0eta_solid_coef}, the coefficients $A_1$, describing solely the $D_{33}$ wave,
 are all in agreement within their error bars. Only in the last energy bin, the coefficients obtained
 for the nuclear targets tend to be slightly larger than those obtained for the free proton. The coefficients $A_2$, describing interference of $D_{33}$ with other waves, are systematically smaller than $A_1$ and generally show a good agreement between results for free proton and nuclear targets.

 One of the sources of systematic uncertainties in the $I^\odot(\Phi)$ results comes from the knowledge of the polarization degree of incident photons $P_{\gamma}$,
 which is typically at the level of a few percent. It includes the stability of the polarization degree of the electron beam $P_{e^-}$
 during the period of data taking, which typically varies within 2\% or even less, and the use of $E_\gamma$ as
 an energy-bin center to calculate $P_{\gamma}$ with Eq.~(\ref{pgamma}). The magnitude of $P_{\gamma}$
 varies by 12\%, 8\%, and 4\% over the first, second, and the third energy bin, respectively.
 For the present measurement, such systematic uncertainties are significantly
 smaller than the uncertainties based on the experimental statistics.
 Another systematic uncertainty is from the limited resolution in the angle $\Phi$,
 which was investigated by comparing the $I^\odot(\Phi)$ from the BnGa PWA polarized amplitude with the asymmetry reconstructed from MC events weighted with this model.
 The resolution impact on $I^\odot(\Phi)$, which was quite small for the results on a free proton~\cite{A2_pi0etap_2018}, is slightly stronger for the case of nuclear targets,
 where the $\gamma N$ c.m.\ frame, calculated by assuming zero nucleon momentum,
 could be different from the actual c.m.\ frame because of Fermi motion.
 Also, using the direction of the two-photon system, without constraining
 its invariant mass to the $\eta$ mass, results a poorer angular resolution, compared to the analysis
 on a free proton obtained with the kinematic-fit hypothesis involving the $\eta \to \gamma\gamma$
 constraint. Because the magnitude of this systematic effect was evaluated from the MC simulation
 to be much smaller than the uncertainties coming from fitting experimental $m(\gamma\gamma)$
 distributions, this systematic uncertainty was neglected in the present results.

 The $I^\odot(\Phi)$ results obtained in this work for the nuclear targets and the comparison
 of them with the free proton indicates the similarity in the $\pi^0\eta$ photoproduction
 mechanism, which is dominated by the $D_{33}$ partial wave.
 Such an observation is consistent with previous measurements performed by the A2 Collaboration with
 a deuterium target~\cite{Kaeser2016}, where the results for $I^{\odot}(\Phi)$ on a quasi-free proton
% and neutron were found to be in agreement with the free-proton data.
 and neutron were found to be in agreement with each other and with the free-proton data.
 Assuming then that the mechanism of the $\pi^0\eta$ photoproduction is dominated by the $D_{33}$ partial wave for heavy nuclei, it appears that FSI effects do not affect
 the asymmetry $I^{\odot}(\Phi)$ significantly for the carbon, aluminum, and lead targets, in analogy
 to the deuteron results. Such a feature was predicted by the Mainz model.
 It was also checked that switching off the $D_{33}$ contribution in this model results in the energy dependence and shape of $I^\odot(\Phi)$ that are notably different from the results observed for the nuclear targets and the free proton, which is illustrated in Fig.~\ref{fig:Io_asym_solid_3x3} for the carbon target. This is another strong indication that the contribution from the $D_{33}$ partial wave is still vital in $\pi^0\eta$ photoproduction on heavy nuclei.

 The results of this work motivate for further searches for in-medium modifications of resonances by using polarization observables in general and $I^{\odot}(\Phi)$ in particular. Besides the fact that the polarization observables allow for a complementary approach to the investigation of photoproduction of resonances, in addition to the unpolarized observables, the empirical observations and model calculations indicate that the impact on the polarization observables from FSI could be less than on the unpolarized.

\section{Summary and conclusions}
\label{sec:Conclusion}
 The beam-helicity asymmetry was measured, for the first time, in photoproduction of $\pi^{0}\eta$ pairs
 on the carbon, aluminum, lead nuclei, with the A2 experimental setup at MAMI. The $I^\odot(\Phi)$ data obtained in the $\pi^0\eta$ photoproduction on the three heavy nuclei were compared
 with the free-proton data. The agreement, observed within the statistical uncertainties, indicates
 that the production mechanism is quite similar to the case of a free nucleon and is therefore dominated
 by the $D_{33}$ partial wave with the $\eta\Delta(1232)$ intermediate state.
 Such an observation is consistent with calculations performed within the Mainz model and
 with previous experimental studies on the deuteron.
 The combined consideration of these observations allows an assumption that
 the beam-helicity asymmetry $I^\odot(\Phi)$ is much less affected by FSI, compared to the unpolarized cross
 section, which provides an opportunity for investigating in-medium modifications of baryon resonances
 without strong influence of FSI effects. In a broader consideration, this paper provides the first results for the photoproduction of $\pi^{0}\eta$ pairs on carbon and heavier nuclei, thus opening a route for further studies of this kind.

\section*{Acknowledgment}

 The authors wish to acknowledge the excellent support of the accelerator group and
 operators of MAMI.
 We also thank A. Sarantsev and V. Nikonov on behalf of the BnGa PWA group.
 This work was supported by the Deutsche Forschungsgemeinschaft (SFB443,
 SFB/TR16, and SFB1044), DFG-RFBR (Grant No. 09-02-91330), the European Community-Research
 Infrastructure Activity under the FP6 ``Structuring the European Research Area''
 program (Hadron Physics, Contract No. RII3-CT-2004-506078), Schweizerischer
 Nationalfonds (Contract Nos. 200020-156983, 132799, 121781, 117601, 113511),
 the U.K. Science and Technology Facilities Council (STFC 57071/1, 50727/1),
the U.S. Department of Energy (Offices of Science and Nuclear Physics,
 Award Nos. DE-FG02-99-ER41110, DE-FG02-88ER40415, DE-SC0014323)
 and National Science Foundation (Grant Nos. PHY-1039130, IIA-1358175),
 INFN (Italy), and NSERC (Canada). We also acknowledge the support of the Carl-Zeiss-Stiftung.
 A.~Fix acknowledges additional support from the Tomsk Polytechnic University competitiveness enhancement program.
 We thank the undergraduate students of Mount Allison University
 and The George Washington University for their assistance.


\begin{thebibliography}{00}

\bibitem{PDG} M.~Tanabashi {\it et al.}, (Particle Data Group),
 Phys.\ Rev.\ D\ {\bf 98}, 030001 (2018).

\bibitem{Anisovich2016} A.~V.~Anisovich {\it et al.},
  Eur.\ Phys.\ J.\ A\ {\bf 52}, 284 (2016).

\bibitem{BeckThoma2017} R.~Beck and U.~Thoma,
 EPJ \ Web \ Conf.\ {\bf 134}, 02001 (2017).

\bibitem{CredeRoberts2013} V.~Crede and W.~Roberts,
 Rept.\ Prog.\ Phys.\ {\bf 76}, 076301 (2013).

\bibitem{Tiator2011} L.~Tiator, D.~Drechsel, S.~S.~Kamalov, and M.~Vanderhaeghen,
  Eur.\ Phys.\ J.\ ST\ {\bf 198}, 141 (2011).

\bibitem{KlemptRichard2010} E.~Klempt and J-M.~Richard,
 Rev.\ Mod.\ Phys.\ {\bf 82}, 1095 (2010).

\bibitem{Frommhold1992} Th.~Frommhold {\it et al.},
  Phys.\ Lett.\ B\ {\bf 295}, 28 (1992).

\bibitem{Bianchi1993} N.~Bianchi {\it et al.},
 Phys.\ Lett.\ B\ {\bf 299}, 219 (1993).

\bibitem{Bianchi1994} N.~Bianchi {\it et al.},
 Phys.\ Lett.\ B\ {\bf 325}, 333 (1994).

\bibitem{Krusche2011} B.~Krusche,
 Eur.\ Phys., J.\ ST\ {\bf 198}, 199 (2011).

\bibitem{KruscheReview2012} B.~Krusche, JPCS {\bf 349}, 012003 (2012).

\bibitem{PPN-BK} B.~Krusche, Prog.\ Part.\ Nucl.\ Phys.\ {\bf 55}, 46 (2005).

\bibitem{Krusche2004} B.~Krusche {\it et al.},
 Eur.\ Phys.\ J.\ A\ {\bf 22}, 347 (2004).

\bibitem{PLB-Robig} M.~R{\"o}big-Landau {\it et al.},
 Phys.\ Lett.\ B\ {\bf 373}, 45 (1996).

\bibitem{PLB-Yor} T.~Yorita {\it et al.},
 Phys.\ Lett.\ B\ {\bf 476}, 226 (2000).

\bibitem{NP-Yam} H.~Yamazaki {\it et al.},
 Nucl.\ Phys.\ A {\bf 670}, 202 (2000).

\bibitem{Dieterle2015} M.~Dieterle {\it et al.},
Eur.\ Phys.\ J.\ A\ {\bf 51}, 142 (2015).

\bibitem{Sokhoyan2015EPJA} V.~Sokhoyan {\it et al.},
Eur.\ Phys.\ J.\ A\ {\bf 51}, 95 (2015).

\bibitem{Sokhoyan2015PLB} V.~Sokhoyan {\it et al.},
Phys.\ Lett.\ B\ {\bf 746}, 127 (2015).

\bibitem{Thiel2015} A.~Thiel {\it et al.},
Phys.\ Rev.\ Lett.\ {\bf 114}, 091803 (2015).

\bibitem{Oberle2013} M.~Oberle, {\it et al.},
Phys.\ Lett.\ B\ {\bf 721}, 237 (2013).

\bibitem{Zehr2012} F.~Zehr, {\it et al.}
Eur.\ Phys.\ J.\ A\ {\bf 48}, 98 (2012).

\bibitem{p2pi0mamic} V.~L.~Kashevarov {\it et al.},
 Phys.\ Rev.\ C\ {\bf 85}, 064610 (2012).

\bibitem{Thoma2008} U.~Thoma, {\it et al.}
Phys.\ Lett.\ B\ {\bf 659}, 87 (2008).

\bibitem{Sarantsev2008} A.~V.~Sarantsev {\it et al.},
Phys.\ Lett.\ B\ {\bf 659}, 94 (2008).

\bibitem{Strauch2005} S.~Strauch {\it et al.},
Phys.\ Rev.\ Lett.\ {\bf 95}, 162003 (2005).

\bibitem{Assafiri2003} Y.~Assafiri {\it et al.},
Phys.\ Rev.\ Lett.\ {\bf 90}, 222001 (2003).

\bibitem{Doering2006} M.~D\"{o}ring, E.~Oset, and D.~Strottman,
Phys.\ Rev.\ C\ {\bf 73}, 045209 (2006).

\bibitem{Mainz_2008} A.~Fix, M.~Ostrick, and L.~Tiator,
  Eur.\ Phys.\ J.\ A\ {\bf 36}, 61 (2008).

\bibitem{Mainz_2010} A.~Fix, V.~L.~Kashevarov, A.~Lee, and M.~Ostrick,,
  Phys.\ Rev.\ C\ {\bf 82}, 035207 (2010).

\bibitem{Kashev_pi0eta_prod} V.~L.~Kashevarov {\it et al.},
 Eur.\ Phys.\ J.\ A\ {\bf 42}, 141 (2009).

\bibitem{Kashev_pi0eta_asym} V.~L.~Kashevarov {\it et al.},
 Phys.\ Lett.\ B\ {\bf 693}, 551 (2010).

\bibitem{A2_pi0etap_2018} V.~Sokhoyan {\it et al.},
  Phys.\ Rev.\ C\ {\bf 97}, 055212 (2018).

\bibitem{CBELSA_pi0etap_2008} I.~Horn {\it et al.},
  Eur.\ Phys.\ J.\ A\ {\bf 38}, 173 (2008).

\bibitem{CBELSA_pi0etap_2010} E.~Gutz {\it et al.},
  Phys.\ Lett.\ B\ {\bf 687}, 11 (2010).

\bibitem{CBELSA_pi0etap_2014} E.~Gutz {\it et al.},
  Eur.\ Phys.\ J.\ A\ {\bf 50}, 74 (2014).

\bibitem{GRAAL_2008} J.~Ajaka {\it et al.},
 Phys.\ Rev.\ Lett.\ {\bf 100}, 052003 (2008).

\bibitem{LNS2006} T.~Nakabayashi {\it et al.},
Phys.\ Rev.\ C\ {\bf 74}, 035202 (2006).

\bibitem{Kaeser2016} A.~K\"aser {\it et al.},
 Eur.\ Phys.\ J.\ A\ {\bf 52}, 272 (2016).

\bibitem{Kaeser2018} A.~K\"aser {\it et al.},
 Phys.\ Lett.\ B\ {\bf 786}, 305 (2018).

\bibitem{Mainz_2005} A.~Fix and H.~Arenh\"ovel,
  Eur.\ Phys.\ J.\ A\ {\bf 25}, 115 (2005).

\bibitem{Mainz_2011} A.~Fix and H.~Arenh\"ovel,
  Phys.\ Rev.\ C\ {\bf 83}, 015503 (2011).

\bibitem{Mainz_2013} A.~Fix, V.~L.~Kashevarov, and M.~Ostrick,
  Nucl. Phys.\ A\ {\bf 909}, 1 (2013).

\bibitem{BGPWA}
A.~V.~Anisovich, R.~Beck, E.~Klempt, V.~A.~Nikonov, A.~V.~Sarantsev, and U.~Thoma,
 Eur.\ Phys.\ J.\ A\ {\bf 48}, 15 (2012);
 http://pwa.hiskp.uni-bonn.de/baryon.htm

\bibitem{MAMI} H.~Herminghaus {\it et al.},
         IEEE Trans.\ Nucl.\ Sci.\ {\bf 30}, 3274 (1983).

\bibitem{MAMIC} K.-H.~Kaiser {\it et al.},
  Nucl.\ Instrum.\ Methods\ Phys.\ Res.\ A\ {\bf 593}, 159 (2008).

\bibitem{TAGGER} I.~Anthony {\it et al.},
  Nucl.\ Instrum.\ Methods\ Phys.\ Res.\ A\ {\bf 301}, 230 (1991).

\bibitem{TAGGER1} S.~J.~Hall {\it et al.},
  Nucl.\ Instrum.\ Methods\ Phys.\ Res.\ A\ {\bf 368}, 698 (1996).

\bibitem{TAGGER2} J.~C.~McGeorge {\it et al.},
  Eur.\ Phys.\ J.\ A\ {\bf 37}, 129 (2008).

\bibitem{CB} A.~Starostin {\it et al.},
 Phys.\ Rev.\ C\ {\bf 64}, 055205 (2001).

\bibitem{TAPS} R.~Novotny,
  IEEE Trans.\ Nucl.\ Sci.\ {\bf 38}, 379 (1991).

\bibitem{TAPS2} A.~R.~Gabler {\it et al.},
 Nucl.\ Instrum.\ Methods\ Phys.\ Res.\ A\ {\bf 346}, 168 (1994).

\bibitem{etamamic} E.~F.~McNicoll {\it et al.},
 Phys.\ Rev.\ C\ {\bf 82}, 035208 (2010).

\bibitem{slopemamic}
 S.~Prakhov {\it et al.}, Phys.\ Rev.\ C\ {\bf 79}, 035204 (2009).

\bibitem{Olsen_1959} H.~Olsen and L.~C.~Maximon,
  Phys.\ Rev.\ {\bf 114}, 887 (1959).

\bibitem{PID} D.~Watts, {\it Proceedings of the 11th International
              Conference on Calorimetry in Particle Physics},
             Perugia, Italy, 2004 (World Scientific, Singapore, 2005), p. 560.

\bibitem{Laget} J.~M.~Laget,
 Nucl.\ Phys.\ A {\bf 194}, 81 (1972).

\bibitem{Ryckebusch}
J.~Ryckebusch, W.~Cosyn, S.~Stevens, C.~Casert and J.~Nys,
Phys.\ Lett.\ B {\bf 792}, 21 (2019).

\end{thebibliography}
\end{document}